\documentclass[twocolumn,showpacs,preprintnumbers,amsmath,amssymb,prl]{revtex4-1}

\usepackage{graphicx}
\usepackage{epstopdf}
\usepackage{dcolumn}
\usepackage{bm}
\usepackage{color}
\usepackage{forloop}
\usepackage{amsfonts}

\definecolor{blue}{RGB}{0, 0, 150}
\definecolor{green}{RGB}{0,150,0}
\definecolor{red}{RGB}{200, 0, 0}
\definecolor{black}{RGB}{0, 0, 0}

\renewcommand{\l}{\left}
\renewcommand{\r}{\right}

\begin{document}


\title{ Gene expression dynamics with stochastic bursts: \\exact results for a coarse-grained model}

\author{Yen Ting Lin}
\email{yenting@umich.edu}
\affiliation{Theoretical Physics Division, School of Physics and Astronomy, The University of 
Manchester, UK}
\affiliation{Max Planck Institute for the Physics of Complex Systems, Dresden, Germany}
\author{Charles R.~Doering}
\email{doering@umich.edu}
\affiliation{Center for the Study of Complex Systems, University of Michigan, Ann Arbor, MI 48109-1107 USA}
\affiliation{Department of Mathematics, University of Michigan, Ann Arbor, MI 48109-1043 USA}
\affiliation{Department of Physics, University of Michigan, Ann Arbor, MI 48109-1040 USA}

\date{\today}
\begin{abstract}
We present a theoretical framework to analyze the dynamics of gene expression with stochastic bursts.
Beginning with an individual-based model which fully accounts for the messenger RNA (mRNA) and protein populations, we propose a novel expansion of the master equation for the joint process.
The resulting coarse-grained model reduces the dimensionality of the system, describing only the protein population while fully accounting for the effects of discrete and fluctuating mRNA population.
Closed form expressions for the stationary distribution of the protein population and mean first-passage times of the coarse-grained model are derived and large-scale Monte Carlo simulations show that the analysis accurately describes the individual-based process accounting for mRNA population, in contrast to the failure of commonly proposed diffusion-type models.
\end{abstract}

\maketitle
Intrinsic noise originating from the discreteness of interacting particles plays an important role in genetic expression: it diversifies the distribution of protein population, promotes transition between different cellular phenotypes on a population level, and in turn enhances organisms' ability to adapt to changing environments without the need of genetic mutation \cite{Kaern}.
There are two primary sources of intrinsic noise in the context of gene expression: \emph{transcriptional noise} from the stochastic transition between active and repressed states of DNA transcription, and \emph{translational noise} from the relatively fast action of mRNA to produce the proteins \cite{Kaern,Walczak}.
Both steps result in bursts of protein production which are experimentally observed \cite{Ozbudak, Blake}.

Many stochastic models have been proposed to model gene circuits \cite{Kepler,Hornos,WalczakSasai,Warren,WangHuang,Wang,Assaf,Strasser,Zhou,Lu} but only a few studies quantitatively account for the effects of bursting noise \cite{Thattai,Friedman,Assaf,Strasser}.
To our knowledge, current theoretical investigation of the dynamical properties of such bursting processes is limited to stationary properties of the protein distribution on the population level \cite{Thattai, Friedman}. 

This Letter presents a new mathematical framework to analyze bursting noise in gene expression.
Starting from an individual-based model including both mRNA and protein populations we construct a novel coarse-grained process describing only the protein population dynamics that fully accounts for the discreteness effects and fluctuations in the mRNA population.
When the mRNA degrades at a much shorter time scale, the approximating process converges to currently proposed bursting models \cite{Thattai, Friedman}.
In our process-based framework, mean first-passage times in a autoregulated gene circuit with stochastic bursts can be formulated and calculated.

We present analytic solutions along with computational verification from large-scale Monte-Carlo simulations.
A key conclusion is that the conventional diffusion approximation of the master equation fails to accurately estimate switching times of the individual-based model. 


\begin{figure}
\includegraphics[width=0.45\textwidth]{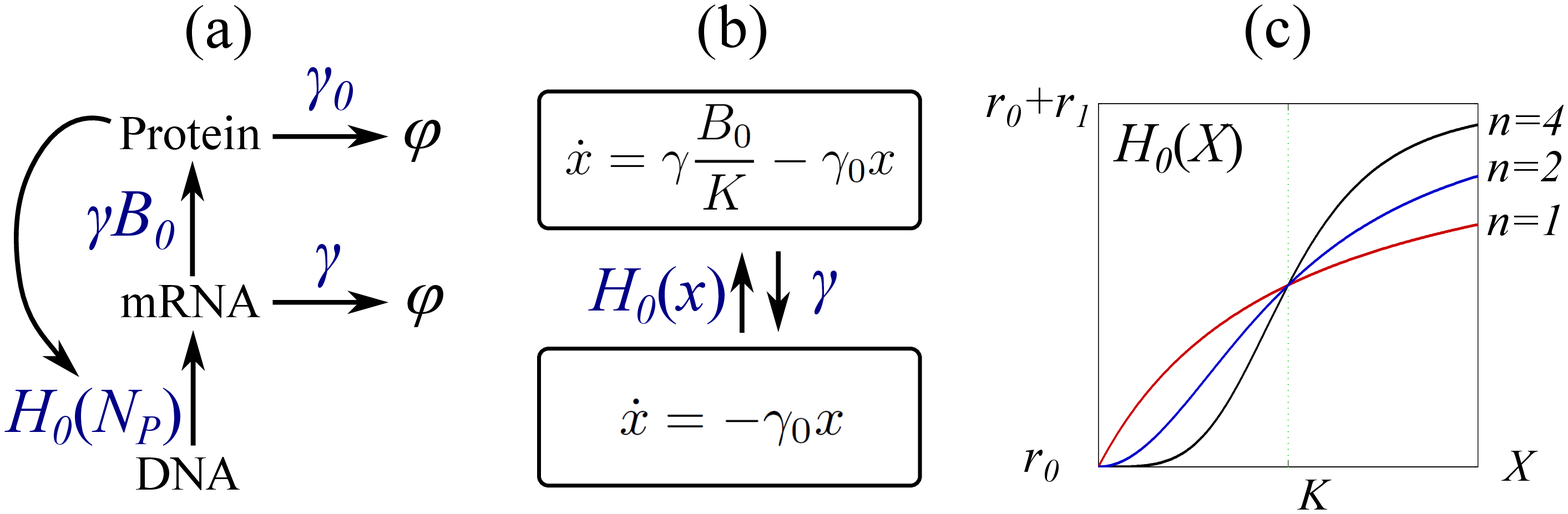}
\caption{Schematic diagrams of (a) the individual-based model and (b) the piecewise deterministic Markov process model \eqref{eq:model}. (c) Hill functions with Hill coeffcients $n=1,2,4$. } \label{fig:1}
\end{figure}

A simple individual-based model of autoregulated gene expression including both the mRNA and protein populations contains four reaction steps \cite{Thattai,Walczak} as summarized in Fig.~\ref{fig:1}(a): synthesis of mRNA's (transcription, $\phi \xrightarrow{H_0\l(N_{\rm P}\r)}{\rm mRNA}$), production of the protein (translation, ${\rm mRNA}  \xrightarrow{\gamma B_0} {\rm mRNA} + {\rm Protein}$), and degradation of both the mRNA's and proteins (${\rm mRNA} \xrightarrow{\gamma} \phi$ and ${\rm Protein}  \xrightarrow{\gamma_0}$).
In the first step $N_{\rm P}$ is the random number of available proteins and in this autoregulated genetic circuit, the population of proteins regulates the transcription rate. The Hill function $H_0(X):= r_0 + r_1 X^n/(K^n + X^n)$ with the Hill coefficient $n$ approximates the autoregulated transcription rate when the gene switches between on and off states on a much shorter time scale \cite{Walczak}.

We refer to the process in Fig.~\ref{fig:1}(a) as the individual-based (IB) model. 
Although the IB model provides a detail description of both the mRNA and protein populations, it is generally difficult to analyze theoretically except for linear cases \cite{Swain,Kumar}.
Single-species models describing only the protein populations are often adopted, especially for more complicated genetic circuits \cite{Wang,Zhou,WangHuang,Lu}.
However, fluctuations in the mRNA population are an important dynamical factor \cite{Assaf,Strasser} and our objective is to construct a coarse-grained model describing only the protein population accounting for contributions from fluctuation in the mRNA population.

Generally mRNA's degrade much faster than proteins. 
In the model organism \emph{Escherichia coli} for example, the mean lifetime of the mRNA is about $2 \min$ while protein lifetimes are $45\sim60 \min$ \cite{Thattai}.
As a consequence, a large number of proteins is produced in a relatively short period of time---a phenomenon termed \emph{translational bursting}.
In addition, due to small system size (the volume of \emph{E.~coli} are $\sim10^{-18} {\rm m}^3$), the onset of the transcription and the lifetime of the synthesized mRNA are observed in a stochastic manner \cite{Cai}.

Motivated by the observation of translational bursting, we propose a novel expansion to approximate the master equation of the IB process in Fig.~\ref{fig:1}(a).
First, we notice that in the IB model, for any given mRNA number $m$, the protein population $N_{P}(t)$ is a birth-death processes with constant birth rate $m \gamma B_0$ and constant per capita death rate $\gamma_0$. 
Therefore, it is convenient to expand the process describing the protein dynamics \emph{conditioning on} the mRNA population: each ``state'' of the system is labeled by the mRNA number $m$.
The transition rate from state $m$ to $m+1$ mRNA molecules is the autoregulated transcription rate $H_0(N_{P})$, and the transition rate from state $m+1$ to $m$ mRNA molecules is the mRNA degradation rate $\gamma$.
Within each state of the system we perform a Kramers--Moyal expansion of the birth-death process \cite{vanKampen,KurtzFP} with respect to the system size $K \gg 1$.
In the lowest order approximation only the advection terms describing the mean-field dynamics are retained \cite{KurtzInf}.
Formally letting the protein concentration be $x:=N_\text{P} / K\ge0$ and the number of mRNA molecules be $m \in \l\{0,1,2,\ldots\r\}$, in each state the protein density evolves according to the deterministic equation
\begin{equation}
\dot{x}(t) = m \gamma \frac{B_0}{K} - \gamma_0 x,
\end{equation} 
with transition rates between different states
\begin{equation}
m \xrightarrow{H(x)} m+1 \text{ and } m \xrightarrow{\gamma} m-1
\end{equation}
where $H(x):=H_0(Kx)$ is the scaled Hill-function.

Next we note that the mean lifetime of the mRNA is $\mathcal{O} (1/\gamma)$ and in the fast-degrading mRNA limit $\gamma \gg 1$, most of the time the system has either $m=0$ or $m =1$.
We therefore neglect states $m \ge 2$ and formulate a closed forward equation for $p_m(x,t)$, the joint probability density that the system presents $m\in \l\{0,1\r\}$ mRNA molecules and protein density $x$ at time $t$:
\begin{equation}
\frac{\partial}{\partial t} \left(\begin{array}{c} p_1(x,t)\\ p_0(x,t)\\ \end{array} \right) =
L^{\dagger}
\left(\begin{array}{cc} p_1(x,t)\\ p_0(x,t)\\ \end{array} \right), \label{eq:model}
\end{equation}
where the forward operator \footnote{Notation: the differential operator $\partial_x$ acts as a total differential including the probability distributions $p_{m}(x,t)$ outside the matrix.} is defined to be
\begin{equation}
L^{\dagger} := \left(\begin{array}{cc} - \gamma - \partial_x \l(\gamma b - \gamma_0 x\r)  & 
H(x) \\ \gamma & -H(x)    + \gamma_0 \partial_x x  \\ \end{array} \right) \label{eq:forward op}
\end{equation}
where we have defined $b := B_0 / K$ to be a dimensionless parameter characterizing the strength of the bursts.
We shall refer to \eqref{eq:model} as the piecewise deterministic Markov process (PDMP; schematic diagram Fig.~\ref{fig:1}(b)) and remark that the process in $x$ alone is non-Markovian. 

\begin{figure}
\includegraphics[width=.48\textwidth]{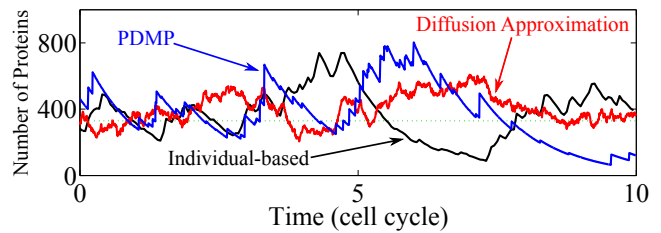}
\caption{Sample paths of the models. Dotted green line denotes 165 molecules which separates low and high protein abundance mode. }\label{fig:2}
\end{figure}

To proceed with our analytic investigation, an infinitely fast-degrading mRNA limit $\gamma \rightarrow \infty$ is taken. Although in such a limit the mean duration when the system stays in $m=1$ state is $1/\gamma \rightarrow 0$, the protein concentration in $m=1$ state increases with a rate $b \gamma \rightarrow \infty$, preserving exponentially distributed random burst with an average burst strength $b$.
In this limit the process stays in $m=0$ state almost surely (i.e.~$p_1(x,t) \rightarrow 0\  \forall t$), and the probability distribution $p_0$ satisfies a closed and second-order differential equation
\begin{equation}
\left( 1 + b \partial_x \right) \partial_t p_0 = 
- \partial_x \left[-x +  b H(x) - b \gamma_0 \partial_x x \right] p_0.\label{eq:FE}
\end{equation}
The stationary probability distribution is obtained by direct integration:
\begin{equation}
p_{\rm stat}(x) = \frac{{\cal N}}{\gamma_0 x} \, \exp\left\{\frac{-x}{b} +  \int^x{\frac{H(\xi)d\xi}{\gamma_0 \xi}}\right\}
\end{equation}
where $\cal N$ is the normalization factor.
Substituting the explicit form of the Hill function we find the analytic expression for the stationary distribution
\begin{equation}
p_{\rm stat}(x) = \frac{\cal N}{\gamma_0} e^{-\frac{x}{b} }x^{ \frac{r_0}{\gamma_0} - 1}  \l(x^{n} + 1 ^{n} \r)^{\frac{r_1}{ n \gamma_0 }}. \label{eq:stationary}
\end{equation}
We remark that in the limit $\gamma \rightarrow \infty$ the PDMP model reduces to the bursting model described by a continuous master equation, and that this result confirms \cite{Friedman}.

In some parameter regimes the stationary distribution \eqref{eq:stationary} exhibits bi-stability \cite{HLnit} and can be adopted to model a biological switch \cite{Friedman,Assaf}.
Our formulation \eqref{eq:model} can be used to derive the mean switching time (MST) between two modes of gene expression in a straightforward way \cite{D1986}.
We begin by deriving the mean first-exit time to leave a domain $(x_1, x_2)$ where $0 < x_1 <x_2<\infty$.

If the initial protein concentration is $x \in (x_1, x_2)$ and the initial number of mRNA is $m$, then the mean time to exit the domain $T_{m} (x)$ satisfies the inhomogeneous equation \cite{vanKampen}
\begin{equation}
- \left(\begin{array}{cc}1\\ 1\\ \end{array} \right) = L\left(\begin{array}{cc}T_1(x,t)\\ T_0(x,t)\\ \end{array} \right), \label{eq:backward}
\end{equation}
where the generator $L$ is the adjoint of the forward operator in \eqref{eq:forward op},
\begin{equation}
L= \left(\begin{array}{cc} - \gamma + \l( \gamma b- \gamma_0 x\r) \partial_x & 
		\gamma \\ H(x) & -H(x)  - \gamma_0 x \partial_x \\ \end{array} \right)
\end{equation}
with boundary conditions
\begin{equation}
		T_1(x_2) = 0,\, T_0(x_1) = 0.  \label{eq:BC}
\end{equation}
The physical meaning of the boundary conditions is clear: when the system starts with the state $m=1$---a state with fast production of proteins---at upper boundary $x_2$, and when the state $m=0$---a state with only degrading proteins---at lower boundary $x_1$, immediately the flow leaves the domain $(x_1, x_2)$.

\begin{figure}
\includegraphics[width=.45\textwidth]{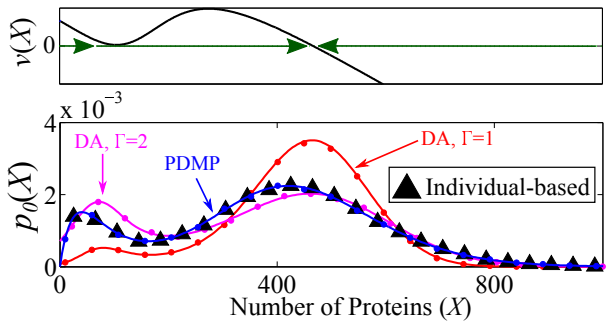}
\caption{Top panel: Drift of the mean-field dynamics $\dot{x} = bH(x) - \gamma_0 x$ showing a single fixed point. Bottom panel: Stationary probability distributions of individual-based model, piecewise deterministic Markov process (PDMP), and the diffusion approximations (DA). Solid lines are analytic solutions. Discrete markers represent numerically measured probability distributions from Monte Carlo simulations.}\label{fig:3}
\end{figure}

Taking the limit $\gamma \rightarrow \infty$ we deduce a closed second-order differential equation for $T_0$,
\begin{equation}
-T''_0  - \l[ \frac{H}{f}-\frac{1}{b} + \frac{H}{x} \l(\frac{x}{H}\r)'\r] T'_0 = \frac{1}{b\gamma_0 x} + \frac{H'}{\gamma_0 x H} \label{eq:DE of T0}
\end{equation}
where prime denote derivative with respect to $x$.
The boundary conditions for \eqref{eq:DE of T0} follow from \eqref{eq:BC}:
\begin{equation}
T_0\l(x_1\r) = 0, 1 =H\l(x_2\r)T_0\l(x_2\r) + \gamma_0 x_2 T_0'\l(x_2\r). \label{eq:BC2}
\end{equation}
We remark that while formally deriving the backward equations of the bursting models \cite{Thattai,Friedman} considering only the protein concentration is possible, imposing the correct boundary conditions \eqref{eq:BC2} is not trivial. 

The solution (derived in the Supplemental Material) is
\begin{equation}
T_0(x)  = C \int_{x_1}^{x} e^{-M(y)} dy + \int_{x_1}^{x} e^{-M(y)} V(y) dy, \label{eq:T- sol}
\end{equation}
where the auxiliary functions $M(x)$ and $V(x)$ and the constant $C$ are
\begin{align}
M(x) :={}& \int^x \l[\frac{H(y)}{f(y)}-\frac{1}{b} + \frac{d}{dy} \l(\ln \frac{y}{H(y)}\r)\r] dy, \\
V(x) :={}&\int^x \l(\frac{-1}{b\gamma_0 y} + \frac{-1}{ \gamma_0 y H(y) }\frac{dH(y)}{dy}\r) e^{M(y)} dy,
\end{align}
\begin{widetext}
\begin{align}
C \equiv \l[-V(x_2) e^{-M(x_2)} f(x_2) - H(x_2) \int_{x_1}^{x_2} V(y) e^{-M(y)} dy + 1\r] \l[f(x_2) e^{-M(x_2)}  + H(x_2)\int_{x_1}^{x_2} e^{-M(y)} dy\r]^{-1}. \label{eq:solution fixed interval}
\end{align}
\end{widetext}
This solution is a generalization of results in \cite{Masoliver,D1987}. 

When the system exhibits bi-modality, the mean switching times between two modes of protein expression can be obtained by taking appropriate limits of \eqref{eq:T- sol}.
First, we define a critical density $x_c$ separating the low and high protein abundance modes, then take $x_1 \rightarrow 0$ and $x_2 \rightarrow x_c$ for the low mode, and $x_1 \rightarrow x_c$ and $x_2 \rightarrow \infty$ for the high mode. 
Careful analysis is needed because \eqref{eq:DE of T0} is singular at $x=0$ (and is presented in the Supplemental Material).
The analytic expressions for the mean switching times (MSTs) are 
\begin{align}
T_{\rm low \rightarrow high} \equiv{}& \int_{0}^{x} e^{-M(y)} V(y) dy + C_2,\label{eq:MST1}\\
T_{\rm high \rightarrow low} \equiv{}& \int_{x_c}^{x} e^{-M(y)} \l[V(y) - V(\infty)\r] dy,\label{eq:MST2}
\end{align}
where the constant $C_2$ is
\begin{equation}
C_2 :=\frac{1-  x_cV(x_c)  e^{-M(x_c)}}{H(x_c)} - \int_0^{x_c} e^{-M(y)} V(y)dy.
\end{equation}

We now turn to the diffusion approximation (DA) of the IB process.
To our knowledge there is no standard way to derive DA models for general bursting kernels.
In the Supplemental Material we present the straightforward Kramers--Moyal expansion \cite{vanKampen, KurtzFP} of the master equation of the IB process in the limit $\gamma \rightarrow \infty$ yielding the It\^{o} stochastic differential equation
\begin{equation}
dX_t = \l[b H(X_t) -\gamma_0 X_t \r] dt + \sqrt{\Gamma b^2 H\l(x\r)} \, dW_t \label{eq:Ito SDE}
\end{equation}
where $X_t$ is the random population density of the proteins, $W_t$ is the standard Wiener process and the scaling factor $\Gamma=2$.
An alternative and phenomenological construction the diffusion approximation is to insert the mean and variance of the bursting kernel in the individual-based process (see Supplemental Material) which yields \eqref{eq:Ito SDE} with the scaling factor $\Gamma=1$.
To avoid leaking probabilities to negative densities, we put a reflective boundary at the origin $x=0$. 
Analytic expressions for the stationary distribution and the mean switching times of the diffusion equation are derived by standard analysis \cite{vanKampen} and presented in the Supplemental Material.

We performed numerical simulations to measure the stationary distributions and the mean switching times (MST) in all three models to verify the theoretical analysis.
For the IB model, exact sample paths are generated by standard continuous time Markov chain simulations \cite{Schwartz}.
For the PDMP model, kinetic Monte Carlo simulations can be constructed by generating exact random waiting times to the next transition events \footnote{The waiting times can be generated by deriving the survival functions of each state \cite{Masoliver} and utilizing the inverse transform sampling.}.
In the limit $\gamma\rightarrow \infty$, we adopted a previously proposed algorithm \cite{Bokes}. For the diffusion approximations we construct a standard Euler--Maruyama integrator of \eqref{eq:Ito SDE}.

The parameters were chosen to be in a biologically relevant regime \cite{Friedman,Thattai,Taniguchi}: $K=200$, $n=4$, $B_0=40$, $r_0=2$, $r_1=10$, $\gamma=30$, while $\gamma_0:=1$ is chosen to normalize the unit of the time by a natural cell cycle. 

Fig.~\ref{fig:3} presents the stationary probability distributions of the IB, PDMP, and DA models. Note that the low-mode is noise induced and does not exist in the mean-field dynamics (top panel of Fig~\ref{fig:3}). While the PDMP model captures the stationary distribution of the IB model extremely well, directly expanding the IB stochastic bursting model by Kramers--Moyal expansion (DA with $\Gamma=2$) qualitatively captures the stationary distribution, and the phenomenological DA model with $\Gamma =1$ failed to capture the stability of the low mode.

\begin{figure}
\includegraphics[width=.48\textwidth]{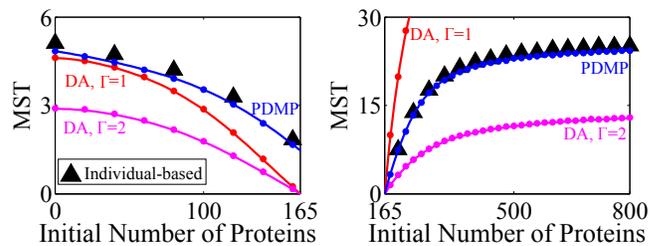}
\caption{Mean switching times (MST) of the individual-based model, piecewise deterministic Markov process (PDMP), and the diffusion approximations (DA). Solid lines are analytic predictions, and discrete markers are measured from Monte Carlo simulations. Left panel: $T_{\rm low \rightarrow high}$; Right panel: $T_{\rm high \rightarrow low}$. $x_c:=165/K$.}\label{fig:4}
\end{figure}

Fig.~\ref{fig:4} presents the MST between low and high protein-abundance modes in all three models. Again, the PDMP model well estimates the mean switching times of the IB model, and both the DA models fail by a large amount.
When the state is initially below $x_c$, both the DA models under-estimate the transition time because the bursting kernels of the DA model have a thinner (Gaussian) tail compared to to the geometric (for the derivation see Supplemental Material) bursting kernel of the IB model.
When the initial state is above $x_c$, the DA model with $\Gamma=1$ over-estimate the MST because the the approximation does not capture the high probabilities of low-density bursts, and the DA model with $\Gamma=2$ underestimate the MST because the approximation fail to capture that the bursting kernel is always positive. 

The PDMP approximation works well even for models with a strong noise strength. In our example, the low-mode is of an order of $100$ protein molecules, and the noise strength (per each burst) is of order $40$ protein molecules.
In addition, the PDMP approximation performs well even though an infinitely-fast degrading mRNA limit $\gamma\rightarrow \infty$ is taken and consequently almost surely there is no mRNA presented in the system.
Meanwhile we observe an average $0.3188$ mRNA in the stationary state of the IB model.


The PDMP model can be easily generalized. For example, finite population and lifetime of mRNA can be considered by generalize \eqref{eq:model} to include $p_m$ with $m\in\l\{0,1,2,\ldots\r\}$.  The transcriptional bursting can be included by considering multiple stages of the gene. Higher dimensional genetic circuit can be investigated by including more states of the system \cite{LinGalla}. These generalizations merit future investigations. 

We conclude that bursting originating from the discreteness of the fast-living mRNA molecules and the stochastic transcription events is the dominating noise in individual-based autoregulated gene expression model.
Diffusion approximations are no longer adequate to analyze the dynamical properties of bursting systems while the novel expansion described here faithfully captures the dynamical properties of the individual-based model in a biologically realistic parameter regime and serves as a new analytic tool to investigate more complex models with bursting noise. 

{\it Acknowledgement.}
YTL was supported by the visitor program at MPI-PKS and EPSRC (grant reference EP/K037145/1).
CRD was supported in part by US-NSF Awards PHY-1205219 and DMS-1515161, and as Simons (Foundation) Fellow in Theoretical Physics.

%

%

\end{document}


\title{ Supplemental Material of \\{ Gene expression dynamics with stochastic bursts: \\exact results for a coarse-grained model}
}

\author{Yen Ting Lin}
\email{yenting@umich.edu}
\affiliation{Theoretical Physics Division, School of Physics and Astronomy, The University of 
Manchester, UK}
\affiliation{Max Planck Institute for the Physics of Complex Systems, Germany}
\author{Charles R.~Doering}
\email{doering@umich.edu}
\affiliation{Department of Mathematics, University of Michigan, USA}
\affiliation{Department of Physics, University of Michigan, USA}
\affiliation{Center for the Study of Complex Systems, University of Michigan, USA}

\maketitle

\section{Individual-Based Model Including the mRNA Population}
In this section, we define the individual-based model including the mRNA populations. The kinetic scheme of the autoregulated process is \cite{Walczak,Thattai}
\begin{equation}
\begin{array}{rll}
\phi {\longrightarrow} & 1\times {\rm mRNA}  & \text{ with a rate }  H_0\l(N_{\rm Protein}\r),\\
{\rm mRNA}  {\longrightarrow}  &  {\rm mRNA}   + 1\times {\rm Protein}  &  \text{ with a rate }  \gamma B_0,\\
{\rm mRNA}  {\longrightarrow}&  \phi&  \text{ with a rate } \gamma ,\\
   {\rm Protein}  {\longrightarrow}&   \phi {}& \text{ with a rate } \gamma_0,\\ 
\end{array}\label{eq:ind-process}
\end{equation}
where the Hill function is defined to be
\begin{equation}
H_0(x) := r_0 + r_1 \frac{x^n}{k^n + x^n}.
\end{equation}
We will denote the (random) population of mRNA and protein by $N_{\rm m}$ and $N_{\rm P}$ respectively.

Define the joint probability distribution of the system at time $t$ to be 
\eq{
P_{m,n}(t):= \mathbb{P} \l\{N_{\rm m}=m , N_{\rm P}=n \text{ at time } t\r\}.
}{}
The master equation of process \eqref{eq:ind-process} is
\begin{align}
\dot{P}_{m,n} ={}& - \l[H_0(n) + \gamma B_0 m+ \gamma m + \gamma_0 n \r]P_{m,n} \\
{}&+  \gamma \l(m+1\r)  P_{m+1,n} + H_0(n) P_{m-1,n} + \gamma b m P_{m,n-1} + \gamma_0 \l(n+1\r) P_{m,n+1} . \label{eq:master eq}
\end{align}
Continuous time Markov chain simulations \cite{Schwartz} are constructed to generate exact sample paths. 

\section{Deriving the PDMP in the large population limit}
In the fast degrading mRNA limit ($\gamma/\gamma_0 \gg 1$), the system only presents only $1$ or $0$ mRNA in a majority portion of the time. Our proposed approximation is to consider the process \eqref{eq:ind-process} conditioning on whether or not the system presents an mRNA, and truncate the probabilities associated with mRNA number greater than 1:
\begin{subequations}
\begin{align} 
P_{m, n}={}&0, \  \forall m>1, \\
\dot{P}_{1,n} ={}& - \l[ \gamma B_0 + \gamma + \gamma_0 n \r]P_{1,n} +  H_0(n) P_{0,n} + \gamma B_0 P_{1,n-1} + \gamma_0 \l(n+1\r) P_{1,n+1},\\
\dot{P}_{0,n} ={}& - \l[H_0(n) + \gamma_0 n \r]P_{0,n} +  \gamma   P_{1,n} +  \gamma_0 \l(n+1\r) P_{1,n+1}.
\end{align}
\end{subequations}

Next, for each of the master equations of the protein number $n$ conditioning on the mRNA number $m$, we perform the conventional Kramers--Moyal expansion\cite{vanKampen}. Denote a typical population scale of the protein by $N_\Omega \gg 1$. Note that in the autocorrelated circuit, it is convenient to choose $N_\Omega=K$. In the continuum limit, the population density is defined by $x := n / K$, and the mean ``burst'' size is defined is $b := B_0 / N_\Omega$. The evolution of the probability distributions $p_0(x,t):=P_{0, n}(t) / K$ and $p_1(x,t):=P_{1, n}(t) / K$  is well-approximated by two coupled Fokker--Planck equations \cite{KurtzFP}
\begin{subequations}
\begin{align}
\partial_t {p}_+={}& -  \gamma {p}_+ +  H(x) {p}_- + \l[\partial_x  \l( \gamma_0 x -\gamma b\r) + \frac{1}{K} \partial_x^2 \l( \gamma_0 x +\gamma b\r) \r] {p}_+,  \label{eq:6a} \\
\partial_t {p}_- ={}& - H(x) {p}_-  +  \gamma  {p}_+ +  \gamma_0\l( \partial_x x+ \frac{1}{2K} \partial_x^2 x\r) {p}_-. \label{eq:6b}
\end{align}\label{eq:6}
\end{subequations}
The coupled Fokker--Planck equation can be expressed in a compact matrix form:
\begin{equation}
\partial_t \left(\begin{array}{cc} p_1\\ p_0 \\ \end{array} \right) =  L^{\dagger}\left(\begin{array}{cc} p_1\\ p_0 \\ \end{array} \right) , \label{eq:piecewise-diffusion}
\end{equation}
with
\begin{equation}
L^{\dagger} := \left(\begin{array}{cc} - \gamma +  \partial_x  \l( \gamma_0 x -\gamma b\r) + \frac{1}{2K} \partial_x^2 \l( \gamma_0 x +\gamma b\r)    & H \\ \gamma & -H  + \gamma_0 \partial_x x+ \frac{\gamma_0}{2NK} \partial_x^2 x\\ \end{array} \right).
\end{equation}
We again remind the reader that the differential operators $\partial_x$ and $\partial_x^2$ act on $p_{0,1}$ too.

It should be clear that the the \emph{discrete population} of the \emph{proteins} causes the demographic stochasticity, which is described by those terms with a prefactor $1/K$. We further propose to take the limit $K \rightarrow \infty$ \cite{KurtzFP} and leave only the advection terms in \eqref{eq:6} to consider exclusively the {\bf bursting noise}, which is a result of \emph{randomly production and degradation events of the mRNA}. In such a limit, the process becomes a piecewise deterministic Markov process: in each state of $m=0$ or $m=1$, the process is deterministic but the switching between the states is Markovian. We emphasize that, in such a limit, the demographic noise which comes from the discreteness of the protein population does not exist---condition on a $m$ state, the concentration of the protein on its own is always evolving in a deterministic fashion.  

We notice that the duration of $m=1$ state is of order $\mathcal{O}\l(1/\gamma\r)$, but the resident time of the $m=0$ state does not depend on $\gamma$---it is of of order $\mathcal{O}\l(\gamma^0\r)$. As a consequence, $p_1$ scales $\mathcal{O}\l(1/\gamma\r)$ and as $\gamma \rightarrow \infty$. In fact, it can be rigorously proved that $p_1 \rightarrow 0$ as $\gamma \rightarrow \infty$, and $\gamma p_1$ in \eqref{eq:6b} can be eliminated by \eqref{eq:6a}:

\begin{align}
\l[1 + b \partial_x  - \frac{ b}{2K} \r]\partial_t p_- ={}&  \l(- b \partial_x  + \frac{ b}{2K} \partial_x^2 \r)  \l[H(x) {p}_-\r] +\gamma_0 \l(1 + b \partial_x  \r) \l[\l( \partial_x x+ \frac{1}{2K} \partial_x^2 x\r) {p}_- \r].\label{eq:diffusion DE}
\end{align}

\section{Individual-Based Bursting Model}
Back to process \eqref{eq:ind-process}, when $\gamma \gg \gamma_0$ and $H(N_{\rm Protein})$, there is a time-scale separation and the mRNA's degrade at a very rapid rate. As a consequence, when one mRNA is formed, almost surely the next happening events before its final degradation are the even more rapid productions of proteins.

Due to the time-scale separation, the production of other mRNA's is and the protein degradation are negligible in one mRNA's lifetime. In such a limit, the distribution of the total number of proteins an mRNA could ever synthesized before its final degradation can be computed. Define the total number of proteins an mRNA could ever synthesize to be $N_\Sigma$, a non-negative random variable.  Because the mRNA has only have two choices---either to degrade or to produce a protein---at any time before the final degradation, the probability that the mRNA produce a protein is $B_0/(1+B_0)$ from reading off the ratio of the rates in the process \eqref{eq:ind-process}.  Therefore, the distribution of $N_\Sigma$ is a geometric distribution
\begin{equation}
\mathbb{P}\l\{N_\Sigma=n\r\}\equiv \l(\frac{B_0}{1+B_0}\r)^{n}\l(\frac{1}{1+B_0}\r) \iff N_\Sigma + 1 \sim {\rm Geom}\l( \frac{B_0}{1+B_0}\r).
\end{equation}
As a consequence, process \eqref{eq:ind-process} in the limit $\gamma \rightarrow \infty$ can be re-formulated to neglect the mRNA population
\begin{equation}
\begin{array}{rll}
\phi {\longrightarrow} & N_\Sigma \times {\rm Protein}  & \text{ with a rate }  H_0\l(N_{\rm P}\r),\\
{\rm Protein}  {\longrightarrow}&   \phi {}& \text{ with a rate } \gamma_0,\\
N_\Sigma +1 \sim{}& {\rm Geom}\l({B_0}/\l({1+B_0}\r)\r).
\end{array}\label{eq:ind-process2}
\end{equation}
We remind the reader that the parameter $B_0$ is the mean number of the proteins an mRNA can produce. We shall refer to model \eqref{eq:ind-process2} as the ``individual-based bursting model''.

Let $P_n$ to be the probability when the system has exactly $n$ proteins. The master equation of process \eqref{eq:ind-process2} can be derived
\begin{equation}
\dot{P_n} = - \l[H_0(n)+\gamma_0 n\r] P_n + \sum_{m=0}^{n} H_0(m) \l(\frac{B_0}{1+B_0}\r)^{n-m}\frac{1}{1+B_0}  P_m+  \gamma_0 \l(n+1\r) P_{n+1}. \label{eq:bursting-master}
\end{equation}

\section{Equivalence between the PDMP and continuous state bursting models}
We now apply Kramers-Moyal system-size expansion is performed \emph{only to} the degradation dynamics in \eqref{eq:bursting-master}. The expansion of \eqref{eq:bursting-master} yields 
\begin{equation}
\partial_t p(x,t) = \l[\gamma_0 \partial_x x + \frac{\gamma_0}{2K} \partial_{x}^2 x\r] p(x,t) + \int_{0}^{x} W(x-y) H(y) p(y) dy,   \label{eq:integro-differential}
\end{equation}
where $p(x,t):=P_n(t) / N_\Omega$ is the continuum probability distribution, $x:=n/N_\Omega$ is the population density of the protein, and $W(x-y)$ is a kernel of the bursting process, defined by the approximating the discrete by the trapezoid rule:
\begin{align}
\int_0^x W(x-y) f(y) dy :={}& -f(x) + \frac{1}{2}\l(\frac{f(0)}{1+ bK} + \frac{f(x)}{1+bK}\r) \nonumber \\
{}&+ \int_0^{x} \frac{1}{1/K+b } e^{-b  N_\Omega \l(x-y\r) \log \l(1+ \frac{1}{bN_\Omega}\r) }f(y) dy.
\end{align}
In the infinity population limit $K \rightarrow \infty$, \eqref{eq:integro-differential} reduces to
\begin{equation}
\partial_t p(x,t) = \partial_x \l[x  p(x,t)\r] + \int_{0}^{x} \frac{e^{-\frac{x-y}{b}}}{b} H(y) p(y) dy -H(x)p(x),  \label{eq:Friedman}
\end{equation} 
which is exactly the continuous master equation in Friedman {\it{et al.}} \cite{Friedman} 

It is straightforward to establish the equivalence of the piecewise deterministic Markov process [i.e., \eqref{eq:6} in the limit $K \rightarrow \infty$] and the continuous-state bursting model \eqref{eq:ind-process2}: acting an operator $1+b\partial_x$ to \eqref{eq:Friedman} yields \eqref{eq:diffusion DE} as $K\rightarrow \infty$.

\section{Deriving the diffusion approximation}
Often in higher dimensional systems, diffusion processes are adopted to analyze complex genetic circuits \cite{Wang,Zhou,WangHuang,Lu}. This section present the derivation to the diffusion approximation of the process \eqref{eq:ind-process2}. 

In the fast-degrading mRNA limit [i.e. $\gamma \rightarrow \infty$ in \eqref{eq:ind-process}], diffusion approximation can be obtained by by performing Kramers--Moyal expansion to \eqref{eq:ind-process2}. The corresponding Fokker--Planck approximation reads
\begin{equation}
\partial_t p_n(x) = - \partial_x\l[ \l(H(x) \frac{\E{N_\Sigma}}{K} -\gamma_0 x \r) p_n\r] + \frac{1}{2 } \partial_x^2 \l[\l(H\l(x\r) \frac{\E{\l[N_\Sigma^2\r]}}{K^2} + \frac{\gamma_0 x}{K^2} \r) p_n\r]
\end{equation}
$N_\Sigma$ is a geometric distribution, and the exact expression of the first two moments are
\begin{subequations}
\begin{align}
\E{N_\Sigma} = {}& B_0\\
\E{N_\Sigma^2} = {}& B_0 \l(1+2B_0\r).
\end{align}
\end{subequations}
When the bursting number is large $B_0 \gg 1$ (typical biological value $10^1 \sim 10^2$ in \emph{E.~Coli} \cite{Thattai,Taniguchi}), we arrive at the final diffusion approximation of \eqref{eq:ind-process2}:
\begin{equation}
\partial_t p_n(x) = - \partial_x\l[ \l(b H(x) -\gamma_0 x \r) p_n\r] + \frac{1}{2 } \partial_x^2 \l[\l(2 b^2 H\l(x\r) + \frac{\gamma_0 x}{K} \r) p_n\r]. \label{eq:SDE}
\end{equation}
We finally remark that in the large population limit, $b$ scales $\mathcal{O}\l(K^{-1}\r)$. A sensible population scaling suggests that near the mean-field fixed points, $\mathcal{O} \l( b H(x) \r) = \mathcal{O} \l(\gamma_0 x \r) = \mathcal{O} \l(K^0\r)$, which in turn indicates that $\mathcal{O} \l( H(x) \r) = \mathcal{O} \l(K\r)$. It is clear that the diffusion term can then be simplified if we \emph{neglect demographic noise} due to the protein degradation, when $B_0 \gg 1$.
\begin{equation}
\partial_t p_n(x) = - \partial_x\l[ \l(b H(x) -\gamma_0 x \r) p_n\r] +  \partial_x^2 \l[\l(b^2 H\l(x\r) \r) p_n\r], \label{eq:SDE2}
\end{equation}
or equivalently the It\^{o} stochastic differential equation
\begin{equation}
dX_t = \l[b H(X_t) -\gamma_0 X_t \r] dt + \sqrt{2 b^2 H\l(x\r)} dW_t, \label{eq:Ito SDE}
\end{equation}
where $dW_t$ is the Wiener process.

In the main text, we refer the diffusion process \eqref{eq:SDE2} to be the diffusion approximation of process \eqref{eq:ind-process} in the fast-degrading mRNA limit ($\gamma \rightarrow \infty$), with a reflective boundary at $x=0$. 

A more phenomenological way is to assert the drift and the diffusion terms of the stochastic differential equation to be the mean ($b$) and variance ($b^2$) of the exponentially distributed burst size. It can be shown \cite{LinGalla} that this approach corresponds to a \emph{constant-burst} model. In this case, the stochastic differential equation is clearly
\begin{equation}
dX_t = \l[b H(X_t) -\gamma_0 X_t \r] dt + \sqrt{b^2 H\l(x\r)} dW_t \label{eq:Ito SDE2}
\end{equation}

Finally, we remark that \eqref{eq:SDE2} is derived from expanding the master equation \eqref{eq:ind-process2} where effect that the bursting noise only enhance the population of the proteins, and \eqref{eq:Ito SDE} clearly over-estimate the noise. On the other hand, the phenomenological approach \eqref{eq:Ito SDE2} clearly under-estimate the low-density bursts of the exponentially distributed kernel.  

\section{Mean switching time of the piecewise deterministic Markov process}
On the domain $\Omega := \l\{x: 0 < x_1 \le x \le x < \infty \r\}$, the backward equation reads
\begin{equation}
- \left(\begin{array}{cc}1\\ 1\\ \end{array} \right) =  \left(\begin{array}{cc} - \gamma + \l[b \gamma -f(x)\r] \partial_x & 
		\gamma \\ H(x) & -H(x)  - f(x) \partial_x \\ \end{array} \right) \left(\begin{array}{cc}T_1(x,t)\\ T_0(x,t)\\ \end{array} \right), 
\end{equation}
and in the limit with fast-degrading mRNA $\gamma \rightarrow \infty$, it converges to
\begin{equation}
- \left(\begin{array}{cc}0\\ 1\\ \end{array} \right) =  \left(\begin{array}{cc} - 1 + b \partial_x & 
		1 \\ H(x) & -H(x)  - f(x) \partial_x \\ \end{array} \right) \left(\begin{array}{cc}T_1(x,t)\\ T_0(x,t)\\ \end{array} \right), \label{eq:matrix DE}
\end{equation}
where we denote $\gamma_0 x$ by $f(x)$. 
It is elementary to eliminate the variable $T_1$ and obtain
\begin{equation}
\frac{d^2 T_0}{dx^2}  + \l[ \frac{H}{f}-\frac{1}{b} + \frac{H}{x} \frac{d}{dx} \l(\frac{x}{H}\r)\r] \frac{d T_0 }{dx} = -\l(\frac{1}{bf} + \frac{1}{fH}\frac{dH}{dx}\r). \label{eq:DE of T-}
\end{equation}
Define an auxiliary function $M(x)$ and $V(x)$
\begin{align}
M(x) := {}& \int^x \l[\frac{H(x')}{f(x')}-\frac{1}{b} + \frac{H(x')}{x'} \frac{d}{dx'} \l(\frac{x'}{H(x')}\r)\r] dx',\\
V(x) := -{}& \int^x \l(\frac{1}{b f(x')} + \frac{1}{ f(x') H(x') }\frac{dH}{dx}\r) e^{M(x')} dx'.
\end{align}
With the expression $H(x):= r_0 + r_1 x^n / (x^n + k^n)$ and $f(x):=\gamma_0 x$, $M(x)$ has a closed form
\begin{equation}
M(x) := \log \l[ e^{-\frac{x}{b}} x^{\frac{r_0}{\gamma_0}+1} \l(x^n + k^n\r)^{\frac{r_1}{n \gamma_0}}\frac{1}{r_0 + \frac{r_1 x^n}{x^n + k^n}}\r]. \label{eq:expM}
\end{equation}
Now \eqref{eq:DE of T-} can be expressed as
\begin{equation}
\frac{d}{dx} \l( e^M(x) \frac{dT_0}{dx}\r) =  -\l(\frac{1}{b \gamma_0 x} + \frac{1}{\gamma_0 x H(x) }\frac{dH}{dx}\r) e^{M(x)},
\end{equation}
and the formal solution is
\begin{equation}
T_0\l(x\r) = C_0 + C_1 \int_{x_1}^{x} e^{-M(x')} dx' + \int_{x_1}^{x} e^{-M(x')} V(x') dx' \label{eq:solution}
\end{equation}
With two constants of integration $C_0$ and $C_1$. 

Since $T_0(x_1) = 0$, clearly $C_0 = 0$. The second constant $C_1$ can be determined by the second boundary condition 
\begin{equation}
T_1(x_2) = 0 \iff -1 = -H(x_2) T_1(x_2) - f(x_2) \frac{dT_1}{dx}(x_2).
\end{equation}
After some algebra, we arrive at
\begin{equation}
C_1 =\frac{
-V(x_2) e^{-M(x_2)} - \frac{1}{f(x_2)} \l[ H(x_2) \int_{x_1}^{x_2} V(x') e^{-M(x')} dx' - 1\r] 
}{
e^{-M(x_2)}  + \frac{ H(x_2)}{f(x_2)} \int_{x_1}^{x_2} e^{-M(x')} dx'.
}
\end{equation}

For the mean switching times between the high- and the low-concentration mode, we have to impose either $x_1 \rightarrow 0$ (for initial state in the low mode) or $x_2 \rightarrow \infty$ (for the initial state in the high mode).

\subsection{$x_1 \rightarrow 0$ limit}
Note that $\exp \l[-M(x)\r]$ has a singularity at $x=0$. However, $V(x)$ is a well-behave function near $x=0$, and we claim 
\begin{equation}
\int_{0}^{x} e^{-M(y)} V(y) dx' < \infty. \label{eq:bob}
\end{equation}
To show this, first let $0<\epsilon \ll 1$, and
\begin{equation}
\int_{0}^{x} e^{-M(y)} V(y) dy = \int_{0}^{\epsilon} e^{-M(y)} V(y) dy + \int_{\epsilon}^{x} e^{-M(y)} V(y) dy.
\end{equation}
The second term is bounded.  As for the first term, since $y < \epsilon \ll 1$, we have
\begin{align}
e^{-M(y)} ={}&  \frac{e^{\frac{y}{b}}} {y^{\frac{r_0}{\gamma_0}+1} \l(y^n + k^n\r)^{\frac{r_1}{n \gamma_0}}}\l(r_0 + \frac{r_1 y^n}{y^n + k^n}\r) \nonumber \\
 \le{}& \frac{e^{\frac{\epsilon}{b}}\l(r_0 + r_1\r)} { k^{\frac{r_1}{\gamma_0}}} \frac{1}{y^{\frac{r_0}{\gamma_0}+1}} =: \frac{B_1}{y^{\frac{r_0}{\gamma_0}+1}}
\end{align}
with a constant $B_1 < \infty$. Similarly,
\begin{equation}
e^{M(y)} \le y^{\frac{r_0}{\gamma_0}+1}\frac{\l(\epsilon^n + k^n\r)^{\frac{r_1}{n \gamma_0}}}{r_0} =:B_2  y^{\frac{r_0}{\gamma_0}+1}
\end{equation}
with a constant $B_2 < \infty$. Similarly, $V(y)$ can be bounded:
\begin{equation}
V(y)\equiv \int^y \l(\frac{1}{b \gamma_0 z} + \frac{1}{ \gamma_0 z H(z) }\frac{dH(z)}{dz}\r) e^{M(z)} dz   \le  B_3 \int^y  z^{\frac{r_0}{ \gamma_0}}  dz = \frac{B_3}{\frac{r_0}{\gamma_0} +1} y^{\frac{r_0}{ \gamma_0}},
\end{equation}
with some $B_3<\infty$. Finally, we have 
\begin{subequations}
\begin{align}
\int_{0}^{\epsilon} e^{-M(y)} V(y) dy \le {}&= \frac{B_1 B_3}{\frac{r_0}{\gamma_0} +1}\int_{0}^{\epsilon} dy < \infty,
\end{align}
\end{subequations}
which establishes our claim \eqref{eq:bob}. 

Next, we proceed to show the following statements:
\begin{equation}
\lim_{x_1 \rightarrow 0}\l[ \int_{x_1}^{x_2} e^{-M(x)} dx  \r]^{-1}=0, \text{ and } \lim_{x_1 \rightarrow 0}  \frac{\int_{x_1}^{x_2} e^{-M(x)} dx }{\int_{x_1}^{x_3} e^{-M(x)} dx } = 1
\end{equation}
for any $x_3 < x_2$. To show this, again we separate the integral
\begin{equation}
\int_{x_1}^{x} e^{-M(y)} dy = \int_{x_1}^{\epsilon} e^{-M(y)} dy + \int_{\epsilon}^{x} e^{-M(y)}  dy.
\end{equation}
The second term is again bounded, and for simplicity define
\begin{equation}
B_5 :=\int_{\epsilon}^{x} e^{-M(y)}  dy < \infty.
\end{equation} 
As for the first term, we begin with the lower bound of the integrand:
\begin{equation}
e^{-M(x)} \ge   \frac{r_0} {y^{\frac{r_0}{\gamma_0}+1} \l(\epsilon^n + k^n\r)^{\frac{r_1}{n \gamma_0}}} =: B_6 \frac{1}{y^{\frac{r_0}{\gamma_0}+1}}.
\end{equation}
As a consequence,
\begin{equation}
\int_{x_1}^{\epsilon} e^{-M(y)} dy \ge \frac{B_6 \gamma_0}{r_0}  \l(\frac{1}{x_1^{\frac{r_0}{\gamma_0}}}  -\frac{1}{\epsilon^{\frac{r_0}{\gamma_0}}} \r),
\end{equation}
and finally
\begin{equation}
\lim_{x_1 \rightarrow 0}\l[\int_{x_1}^{x_2} e^{-M(y)} dy\r]^{-1} \le \lim_{x_1 \rightarrow 0}\frac{r_0}{B_6 \gamma_0} \frac{x^{\frac{r_0}{\gamma_0}}}{x^{\frac{r_0}{\gamma_0}} + \epsilon^{\frac{r_0}{\gamma_0}}} =0.
\end{equation}

Similarly, it is straightforward to apply the L'H\^{o}pital's law to show
\begin{equation}
\lim_{x_1 \downarrow 0}\frac{\int_{x_1}^{x_2} e^{-M(x)} dx }{\int_{x_1}^{x_3} e^{-M(x)} dx } = 1.
\end{equation}

To sum up, upon taking the limit $x_1 \rightarrow 0$, the solution \eqref{eq:solution}---the mean switching time if the system starts with the low-protein-abundance mode---can be expressed as 
\begin{equation}
T_0(x) \equiv \lim_{x_1 \rightarrow 0} T_0(x) = \int_{0}^{x} e^{-M(x')} V(x') dx' + T_0(0)
\end{equation}
with 
\begin{subequations}
\begin{align}
 T_0(0) ={}& \lim_{x_1 \rightarrow 0} C_1 \int_{x_1}^{x} e^{-M(x')} dx'\\
 = {}&  \lim_{x_1 \rightarrow 0}  \frac{
-V(x_2) e^{-M(x_2)} - \frac{1}{f(x_2)} \l[ H(x_2) \int_{x_1}^{x_2} V(x') e^{-M(x')} dx' - 1\r] 
}{\frac{e^{-M(x_2)}  + \frac{ H(x_2)}{f(x_2)} \int_{x_1}^{x_2} e^{-M(x')} dx'}{\int_{x_1}^{x} e^{-M(x')} dx'}
}\\
= {}&  \frac{
-V(x_2) e^{-M(x_2)} - \frac{1}{f(x_2)} \l[ H(x_2) \int_{x_1}^{x_2} V(x') e^{-M(x')} dx' - 1\r] 
}{\frac{ H(x_2)}{f(x_2)}}\\
= {}& \frac{1}{H(x_2)}\l[1-  x_2V(x_2)  e^{-M(x_2)}\r] - \int_0^{x_2} e^{-M(x')} V(x')dx'
\end{align}
\end{subequations}

\subsection{$x_2 \rightarrow \infty$ limit}
Note that $\exp \l[-M(x)\r]$ has a singularity at $x\rightarrow \infty$. On the other hand, $V(x)$ is again well-behaving. Rewrite
\begin{equation}
C_1 =\frac{
-V(x_2)- \frac{1}{f(x_2)} e^{M(x_2)} \l[ H(x_2) \int_{x_1}^{x_2} V(x') e^{-M(x')} dx' - 1\r] 
}{
1  +e^{M(x_2)} \frac{ H(x_2)}{f(x_2)} \int_{x_1}^{x_2} e^{-M(x')} dx',
}
\end{equation}
and we aim to show that $\lim_{x_2 \rightarrow \infty} C_1 = -\lim_{x\rightarrow \infty}V(x) < \infty$. 

First, we show that for a given and strictly positive $x_1$,
\begin{equation}
\lim_{x_2 \rightarrow \infty }e^{M(x_2)}  \int_{x_1}^{x_2} e^{-M(y)} dy =0.
\end{equation}
Clearly from the exact expression of $M(x_2)$, we have 
\begin{align}
e^{M(x_2)}  \int_{x_1}^{x_2} e^{-M(y)} dy = {}&\int_{x_1}^{x_2} e^{-(x_2 - y)}\frac{x_2^{\frac{r_0}{\gamma_0}+1} \l(x_2^n + k^n\r)^{\frac{r_1}{n \gamma_0}}\frac{1}{r_0 + \frac{r_1 x_2^n}{x_2^n + k^n}}}{y^{\frac{r_0}{\gamma_0}+1} \l(y^n + k^n\r)^{\frac{r_1}{n \gamma_0}}\frac{1}{r_0 + \frac{r_1 y^n}{y^n + k^n}}} dy \\
\le {}& \l(1+\frac{r_1}{r_0}\r) \l(1+\frac{k^n}{x_2^n}\r)^{\frac{r_1}{n \gamma_0}} \int_{x_1}^{x_2} e^{y-x_2} \l(\frac{x}{y}\r)^{\frac{r_0}{\gamma_0} + \frac{r_1}{\gamma_0} + 1}  dy.
\end{align}
For $x_2 \gg x_1$, it is elementary to show that the integrand has a single maxima at $ y_\ast := r_0 / \gamma_0 + r_1 / \gamma_0 + 1<\infty$. As a consequence, 
\begin{subequations}
\begin{align}
e^{M(x_2)}  \int_{x_1}^{x_2} e^{-M(y)} dy \le{}&  \l(1+\frac{r_1}{r_0}\r) \l(1+\frac{k^n}{x_2^n}\r)^{\frac{r_1}{n \gamma_0}} e^{y_\ast-x_2} \l(\frac{x_2}{y_\ast}\r)^{\frac{r_0}{\gamma_0} + \frac{r_1}{\gamma_0} + 1} \int_{x_1}^{x_2} dy \\
= {}&  \l(1+\frac{r_1}{r_0}\r) \l(1+\frac{k^n}{x_2^n}\r)^{\frac{r_1}{n \gamma_0}} e^{y_\ast-x_2} \l(\frac{x_2}{y_\ast}\r)^{\frac{r_0}{\gamma_0} + \frac{r_1}{\gamma_0} + 1} \l(x_2 - x_1 \r),
\end{align}
\end{subequations}
and apparently,
\begin{equation}
\lim_{x_2 \rightarrow \infty }e^{M(x_2)}  \int_{x_1}^{x_2} e^{-M(y)} dy =0. \label{eq:cc}
\end{equation}
Note that $V(x)$ is a strictly decreasing and well-behaved function, so $0 > V(x) > V_{\rm min} := \lim_{x\rightarrow \infty} V(x)$ for $x_1 < x < \infty$. As a consequence, 
\begin{equation}
\lim_{x_2 \rightarrow \infty } \l\vert e^{M(x_2)}  \int_{x_1}^{x_2} V(y) e^{-M(y)} dy  \r\vert \le \l\vert V_{\rm min} \r\vert \lim_{x_2 \rightarrow \infty }e^{M(x_2)}  \int_{x_1}^{x_2}e^{-M(y)} dy = 0.  \label{eq:dd}
\end{equation} 
The above Eqs.~\eqref{eq:cc} and \eqref{eq:dd} suggest that
\begin{equation}
\lim_{x_2 \rightarrow \infty } C_1 \rightarrow \l \vert V_{\rm min} \r\vert,
\end{equation} 
and the final solution of the mean switching time if the system begin with a high-protein-abundance state is
\begin{equation}
T_0\l(x\r) = \l\vert V_{\rm min} \r\vert \int_{x_1}^{x} e^{-M(x')} dx' + \int_{x_1}^{x} e^{-M(x')} V(x') dx'.
\end{equation}

\section{Mean switching time of the diffusion approximation}
The mean first-exit time of general one-dimensional diffusion process was extensively studied in the classic book by van Kampen \cite{vanKampen}. For the reference of the reader, in this section we briefly present the key equations used to evaluate the mean switching time of the process \eqref{eq:Ito SDE}

Given any transition boundary $x_c$, we define the domains of interest $\Omega_L:= \l\{x:x<xc\r\}$ and $\Omega_H:= \l\{x: x>x_c\r\}$, respectively the low and high protein abundance states. For the domain $\Omega_L$, we impose a reflective boundary at $x=0$. We are interested in the mean of the first times when the process $X_t$ leaving the domain $\Omega \in \l\{\Omega_L, \Omega_H\r\}$. 

Denote the random first exit time by $\tau\l(x\r) := \inf \l\{t : X_t \notin \Omega \vert X_0=x \r\}$. The mean first exit time $T(x) := \E{\tau\l(x\r)}$ satisfies the following adjoint equation
\begin{equation}
-1 = v(x) \frac{d}{dx} T(x) + D(x) \frac{d^2}{dx^2} T(x). \label{eq:SDE backward}
\end{equation}
Here, we define the drift $v(x)$ and the diffusion $D(x)$ according to \eqref{eq:Ito SDE}:
\begin{align}
v(x):={}& b H(x) -\gamma_0 x ,\\
D(x):={}& b^2 H\l(x\r).
\end{align}

The general solution of \eqref{eq:SDE backward} is
\begin{equation}
T\l(x\r) = C_0 + C_1 G(x) - H(x) \label{eq:MST SDE sol}
\end{equation}
where $C_0$ anc $C_1$ are constant from the integrations, and the auxiliary functions are defined to be
\begin{align}
\phi(x;x_0) := {}& 2 \int^x_{x_0} \frac{v(y)}{D(y)} dy,\\
G(x;x_0):={}& \int^x_{x_0} e^{-\phi(y)}dy,\\
K(x;x_0):={}&2  \int^x_{x_0} \frac{ e^{\phi(y)}}{D(y)}dy.\\
H(x;x_0):={}& \int^x_{x_0} e^{-\phi(y)} K\l(y\r) dy.
\end{align}

The boundary conditions of \eqref{eq:MST SDE sol} depends on which domain the initial state is on. On the one hand, when the state start from a low protein-abundance state, the boundary conditions are
\begin{align}
\frac{dT}{dx}(0)=0, \text{ and } T(x_c)=0
\end{align}
which result in the mean switching time from low to high protein-abundance to high protein-abundance state
\begin{equation}
T(x<x_c)= \l[\lim_{x_1 \rightarrow \infty}H\l(x_1; x_c\r)\r] G(x; x_0) - H(x; x_0).
\end{equation}
On the other hand, when the state begin with a high protein-abundance state, the boundary conditions are
\begin{align}
\lim_{x\rightarrow \infty}\frac{dT}{dx}(x)=0, \text{ and } T(x_c)=0,
\end{align}
and the mean switching time from high to low protein-abundance state is
\begin{equation}
T(x<x_c)= H\l(x_c; 0\r) - H(x;0).
\end{equation}